\documentstyle[aps,twocolumn]{revtex}

\newcommand{\beq}{\begin{equation}}
\newcommand{\eeq}{\end{equation}}
\newcommand{\beqa}{\begin{eqnarray}}
\newcommand{\eeqa}{\end{eqnarray}}
\def\opone{\leavevmode\hbox{\small1\kern-3.8pt\normalsize1}}
\def\half{{1\over 2}}

\def\<{\langle}
\def\>{\rangle}
\def\up{\uparrow}
\def\down{\downarrow}

\title{Coherent quantum measurement for the direct determination of the degree of polarization 
and Polarization Mode Dispersion compensation}
\author
{N. Gisin\\
{\protect\small\em Group of Applied Physics, University of Geneva, 1211 Geneva 4, Switzerland}\\
}
\date{\today}

\begin{document}

\maketitle

\begin{abstract}
An example of a coherent measurement for the direct evaluation of the degree of
polarization of a single-mode optical beam is presented. It is applied to the case
of great practical importance where depolarization
is caused by polarization mode dispersion. It is demonstrated that coherent measurement 
has the potential of significantly increasing the information gain, compared to standard
incoherent measurements.
\end{abstract}


\section{Introduction}
One of the lessons of quantum information theory \cite{QIPIntro98,physQI}
is that coherent measurements generally
provide more information than incoherent ones. And this is true even if the quantum
systems under investigation are not entangled. Recall that coherent measurements are
represented by operators whose eigenvectors are entangled. Consequently, coherent
measurements, contrary to incoherent ones, necessarily treat several systems at once.
Most known examples use spin $\half$, named in this context {\it qubits}. For instance, to
determine the direction $\vec m$ of 2 spin $\half$ in a pure but unknown state $|\vec m\>\otimes|\vec m\>$,
Massar and Popescu \cite{MassarPopescu1995}
have proven that a specific coherent measurement provides the maximal
possible fidelity \cite{fidelity}, strictly better than all successive standard measurement,
even if the second one is allowed to depend on the result of the first one. 
This is an intellectually attractive result! The game is possibly even
more fascinating if one starts from 2 spin $\half$ in pure, but opposite states 
$|\vec m\>\otimes|-\vec m\>$ \cite{GisinPopescu99}. But, admittedly, all known examples, in particular the
above mentioned ones, are somewhat artificial. Also, in order to have a well defined
and tractable problem, all examples assume simple a priori distributions (e.g. random
directions, but pure state) and use as figure of merit the concept of fidelity \cite{fidelity}, 
whose main quality is to be easy to compute. 

In this letter, on the contrary, 
we consider a problem of great practical value and show that coherent 
measurements could provide a real advantage over all possible incoherent measurements
schemes. Moreover, our proposal can be realized with today's technology and has the
potential of being made very compact and cost effective. The price to pay, conceptually,
is that the figure of merit and the a priori distribution are not uniquely defined. The
concept of {\it optimal} measurement is thus ill-defined.

The problem is to measure the Degree Of Polarization (DOP) of a single mode optical beam,
e.g. at the output of a single-mode optical fiber. This
is usually done indirectly, by measuring all 4 Stokes parameters. This is typically an
incoherent measurement scheme: each photon is measured with a polarizing beam splitter (PBS)
with detectors at each of the outports. For about a third of the photons the PBS 
analyses circular polarization, while for the second and last third of photons it analyses 
the photons in the horizontal-vertical and diagonal linear polarization bases, 
respectively. Such polarimeters are expensive, bulky, rather slow and require 
frequent recalibration. But one can easily convince oneself that there is
no better way to measure the DOP if one treats each photon separately, i.e. if one uses
incoherent measurements. Let us note that treating photons separately does not mean that one
uses single-photon detector (actually, all commercial polarimeters use analog detectors
sensitive to the photon flux); it means that the contribution of each photon is independent
of all the others.

Now, assume we have 2 photons in the same unknown mixed state (we denote polarization states
by vectors, using the Poincar\'e sphere representation): 
\beq
\frac{\opone+\vec M\vec\sigma}{2}\otimes\frac{\opone+\vec M\vec\sigma}{2}
\label{2photonState}
\eeq
where the DOP equals $|\vec M|$, $0\le|\vec M|\le1$, and $\vec\sigma$ denote the Pauli matrices. 
The state (\ref{2photonState}) is
clearly not entangled, but let us consider coherent measurements on such states.
The overlap between
the product state (\ref{2photonState}) and the archetype of entangled state, the singlet
$\psi^{(-)}=\frac{1}{\sqrt{2}}\left(|\up,\down\>-|\down,\up\>\right)$, reads:
\beq
\<\psi^{(-)}|\frac{\opone+\vec M\vec\sigma}{2}\otimes\frac{\opone+\vec M\vec\sigma}{2}
|\psi^{(-)}\> = \frac{1-|\vec M|^2}{4}
\label{Psinglet}
\eeq
Consequently, a coherent measurement with the singlet state as eigenstate provides results
directly related to the DOP. Intuitively this can be understood as follows. If the DOP is high,
the photons are likely to have the same polarization and thus no overlap with the singlet state
(recall that in a singlet state the 2 spin $\half$ are anti-correlated in all direction).

It is natural to ask whether the above coherent measurement is better in principle than a succession of
two incoherent measurements. But, a priori the question seems ill defined because there is no natural
a priori probability distribution for the DOP (for the direction of $\vec M$ we make the
natural assumption of complete randomness, i.e. the direction is uniformly distributed).
One possible distribution for the DOP could be uniform between 0 and 1. Another would
assume that the vector $\vec M$ is uniformly distributed in the sphere equipped with the
euclidean metric. But in fact the a priori distribution depends on the 
problem under consideration. In the next section we present the problem we have in mind
and compute the corresponding probability distribution for the DOP.

\section{Depolarization due to Polarization Mode Dispersion}
Polarization Mode Dispersion (PMD) is presently one of the limiting factor in optical fiber
communication systems (see \cite{PMDimpact} and reference therein). 
In particular, fibers installed in the 1980's have rather large PMDs and there
is an intense effort to develop PMD compensators \cite{PMDcompensation}. 
One difficulty is that PMD is a statistical quantity (see below), the compensator 
must thus be active. Ideally, the feedback parameter
should be the Differential Group Delay (DGD), but the measurement time would be too long. Another
possible feedback parameter is the DOP. There is thus a high interest in fast DOP measurement
techniques. In addition to speed, typically in the microsecond domain, the measurement
should operate at low powers ($\mu$W). In this section we briefly recall the origin of PMD,
the basic concepts of Principal States (PS) and of Differential Group Delay (DGD) \cite{princstate} and 
compute the induced depolarization. This provides the a priori DOP distribution for this
case of great practical importance.

PMD is cause by small asymmetries in the fiber, hence locally there are two group velocities.
This simple picture has to be combined with polarization mode coupling: the degeneracy
between the 2 polarization modes of perfectly circular fibers is only slightly lifted 
by the asymmetry present in all real fibers. Consequently, the 2 polarization
modes couple easily. Accordingly, the photons propagate partly in the fast mode, partly
in the slow mode. The polarization mode coupling is very sensitive to any mechanical or
thermal perturbation. Thus, in
practice, it fluctuates and has to be considered as a statistical phenomenon. If the source
is of very low coherence, the outcoming light is completely depolarized. If, at the
other extreme, the source is monochromatic with optical frequency $\omega$, 
the beam propagating in the single-mode fibre is fully polarized, its
polarization state $\psi_\omega$ resulting from complex interferences. In the realistic cases
the optical spectrum $\Delta\omega$ is finite but small compared to the PMD. One can
then show that there exists two mutually orthogonal 
privilege polarization modes, called the Principal
States (PS) $\psi_\omega^{PS}$, characterized by their stability to first order \cite{princstate}:
\beq
\frac{\partial\psi_\omega^{PS}}{\partial\omega}=\pm\frac{i}{2}\delta\tau(\omega)\psi_\omega^{PS}
\label{psiPS}
\eeq
where $\delta\tau(\omega)$ is the DGD, i.e. the relative delay between the two PS. Note that
(\ref{psiPS}) implies $\frac{\partial}{\partial\omega}\<\vec\sigma\>_{\psi_\omega^{PS}}=0$.
Any incoming
polarization state can thus be decomposed onto the two PS which then propagate without distortion
(neglecting chromatic dispersion), but at slightly different velocities, as illustrated
on Figure 1. Assuming incoming pulses of Gaussian shape,
$I(t)=I_0 e^{-t^2/2\sigma^2}$ with spread $\sigma$, the outcoming pulses in each PS read:
\beqa
E_{fast}(t)=E_0\sqrt{\frac{1+\eta}{2}}e^{(t-\tau_0-\delta\tau)^2/4\sigma^2} \\
E_{slow}(t)=E_0\sqrt{\frac{1-\eta}{2}}e^{(t-\tau_0+\delta\tau)^2/4\sigma^2} 
\eeqa
where $\tau_0$ is the mean propagation time, $E_0$ takes into account the attenuation and 
$\eta$ denotes the relative intensity of each PS ($\eta$ is 
randomly distributed between -1 and 1). 
It is known that for fibers much longer than the
mean polarization mode coupling length (in practice fibers longer than a few km), the
DGD $\delta\tau$ is a random variable with Maxwell distribution 
\cite{Matera90,PMDcoherence,PMDCOST}:
\beq
\rho (\delta\tau ) = \frac{3\delta\tau^2}{\sqrt{\pi /6}~(\Delta \tau)^3} exp{(-\frac{3\delta\tau^2}{2(\Delta \tau)^2})}
\eeq
where the rms DGD $\Delta\tau$ is defined as the PMD.

The polarization state of the outcoming light can now by described by the following density
matrix:
\beqa
\rho&=&\int dt (E_{fast}|fast\>+E_{slow}|slow\>) \\
&& \hspace{1cm} (E_{fast}\<fast|+E_{slow}\<slow|) \\
&=&\frac{\opone+\vec M\vec\sigma}{2}
\eeqa
from which the DOP can be computed:
\beq
DOP=M\equiv|\vec M|=\left(\eta^2+(1-\eta^2)e^{-\delta\tau^2/4\sigma^2}\right)^\half
\eeq
The DOP probability distribution can be computed from:
\beq
\rho(M)=-\frac{d}{dM}~Prob(DOP\ge M)
\label{rhoDOP}
\eeq
where
$Prob(DOP\ge M)=\int_0^\infty d(\delta\tau)\rho(\delta\tau)\int_{\eta_{\min}}^1 d\eta$
and 
$\eta_{\min}^2 = \max\{0,\frac{M^2-e^{-\delta\tau^2/4\sigma^2}}{1-e^{-\delta\tau^2/4\sigma^2}}\}$.
Figure 2 illustrates the corresponding DOP probability distribution for pulses with spread
$\sigma$=10 ps and PMDs $\Delta\tau$=20, 30 and 40 ps.

\section{implementation of the coherent measurement}
Implementation of coherent measurements require 2-photon processes, i.e. non-linear
optics. Efficient 2-photon absorption at telecom wavelength has been demonstrated
using commercial devices \cite{Tsang,Reid}. In these demonstrations, however, the
two photons had to have a fixed polarization, determined by the geometry of the
device. In order to have a selective 2-photon absorption for the singlet state,
interference effects have to be exploited. This is clearly possible using either
gas cells as suggested in \cite{Scully99} for Bell-state analysis, 
or using solid state devices \cite{Tomita}.
More generally, any of the configurations used to produce maximally entangled 
polarization states via parametric downconversion in nonlinear crystals
could be used (see e.g. \cite{InnsbruckSource,KwiatSource}), permuting
the input and output, see figure 3. For the 1550 nm telecom wavelength, the upconverted 
photons around 775 nm can be detected with efficient
Si avalanche photodiodes. Such simple configurations could be improved using cavities 
similar to the techniques used in optical parametric oscillators.

Note that since photons are bosons, two photons can be in the anti-symmetric singlet polarization
state only if either they propagate in different modes \cite{diffmode}, or they have different
wavelengths. When photon pairs are prepared in the singlet state for quantum information 
experiments, they are usually send in different directions. But in the present case they
arrive within the same optical pulse in the same single-mode optical fiber. Hence they can
have a singlet component only if the optical spectrum is not arbitrarily sharp, in
agreement with the fact that a single spatial mode of perfectly monochromatic light is always 100\%
polarized. Accordingly, a possibly convenient collinear configuration to measure the
singlet component, hence to measure the DOP, consists in 2 type II crystals in series
(type II means that the two photons have mutually orthogonal polarizations).
The first crystal's cut should be such that phase matching favors 2-photon absorption 
with the lower frequency photon polarized vertically, while the second crystal is such that
2-photon absorption is favorable when the lower frequency photon is polarized horizontally.
The 2-photon absorptions in the 2 crystals are coherent in the sense that the upconverted photon
results from the coherent superposition of the processes taking place in each crystal. 
The distance between the 
crystals can thus be adjusted such that only photon pairs in the singlet state are upconverted.

\section{Figures of merit}
It is too early to assess the practical value of our proposal, though we plan to work on
this issue. With enough engineering efforts, one can expect a compact device. This
would be the first direct measurement of the DOP, usually measured indirectly via the 4
Stokes parameters. Accordingly, an improved accuracy to measurement-time ratio might
be achieved.

On the conceptual side, let's assume that one has 2 photons in state (\ref{2photonState})
where the direction of $\vec M$ is random and the DOP $|\vec M|$ a priori distribution follows
(\ref{rhoDOP}). If the coherent measurement produces the outcome "singlet", then the
a posteriori distribution reads:
\beq
\rho(M|singlet)=\rho(M)\frac{P(singlet|M)}{P_{singlet}}
\eeq
where $P_{singlet}=\int_0^1 \rho(S)P(singlet|S)dS$ and
$P(singlet|M)$ is given by (\ref{Psinglet}). In this case the information gain
is given by $I_{singlet}\equiv H_0-H_{singlet}$ with the entropies 
$H_0=\int_o^1 dM\rho(M)\log_2(\rho(M))$ and
$H_{singlet}=\int_o^1 dM\rho(M|singlet)\log_2(\rho(M|singlet))$. The case of "triplet"
outcome is similar. The mean information gain for coherent measurement is thus:
$I_{coh.}=P_{singlet}I_{singlet}+P_{triplet}I_{triplet}$. This should be compared to the
information gain using incoherent measurements. The optimal is obtained when measuring
both photons in the same basis. The probability $P(o_1=o_2|M)$ that both photons 
produce the same result $o_j$ is then $P(o_1=o_2|M)=\frac{3+M^2}{6}$. We found numerically that
the information gain using the coherent measurement is always larger than the obtainable
gain using incoherent measurements. For PMDs of 20, 30 and 40 ps, the ratio 
$I_{coh.}/I_{incoh.}$ is 7.08, 5.69 and 5.23, respectively. 
For arbitrary large PMD, the ratio tends to 4.82.

A priori one could imagine that the proposed coherent measurement is optimal (from Shannon's
information point of view) in some cases, but not in others. However, Acin and colleagues \cite{optimal}
recently proved that it is optimal, whatever the a priori distribution is. This is a rather
surprising result obtained in a research on the optimal estimation of 2-qubit entanglement.

\section{Conclusion}
Thanks to recent conceptual and technological progress, quantum physics starts to 
reveal its tremendous power for information processing. 
We presented a scheme based on quantum coherent measurements
which provides direct information on the degree of polarization of a single mode
optical beam. The measurement is feasible using similar nonlinear crystals as used
for photon pair creation via parametric downconversion. The case where depolarization is due
to polarization mode dispersion has been analyzed in detail, including the information 
gain compared to incoherent measurements. This example is of great practical relevance 
for today's fiber optic communication systems.

\small
\section*{Acknowledgements}
Financial support by the Swiss OFES within the European projects COST 265 and
EQUIP (IST-1999-11053) are acknowledged. Help with the figures by Claudio Vinegoni
was greatly appreciated.


\section*{Figure Captions}

\begin{enumerate}
\item{Fig. 1}
Principle of depolarization due to PMD. The incoming pulse (left of the fiber) is decomposed onto the fast
and slow principal polarization states (right of the fiber, see text). 

\item{Fig. 2}
Probability distribution for the DOP in case of Gaussian pulses with a spread of 10 ps and 
optical fibers with PMDs of 20 ps (full line), 30 ps (dahsed) and 40 ps(dotted line).

\item{Fig. 3}
A possible collinear configuration using two nonlinear crystal in series.

\end{enumerate}


\begin{thebibliography}{99}
\bibitem{QIPIntro98} {\it Introduction to Quantum computation and
information}, eds H.K. Lo, S. Popescu \& T.P. Spiller, (World Scientific, 1998)

\bibitem{physQI} {\it The Physics of Quantum information}, Eds D. Bouwmeester, 
       A. Ekert and A. Zeilinger, Springer-Verlag 2000.

\bibitem{MassarPopescu1995} S. Massar and S. Popescu, Phys. Rev. Lett. 
       {\bf74}, 1259, 1995.

\bibitem{fidelity} Overlap between the guessed state and the real one.

\bibitem{GisinPopescu99} N. Gisin and S. Popescu, Phys. Rev. Lett.
       {\bf82}, 432-435, 1999.

\bibitem{PMDimpact} F. Bruy\`ere and O. Audouin, IEEE Photon. Tech. Lett. {\bf6}, 443, 1994;
       M. Karlsson, Optics Lett. {\bf23}, 688, 1998.

\bibitem{PMDcompensation} R. No\'e et al. IEEE J. Lightwave Tech. {\bf17}, 1602, 1999.
       C. Francia et al., Electron. Lett. {\bf35}, 414, 1999. M.W. Chat et al., Proceedings of
       the Conference on Optical Fiber Communication OFC'99, Postdeadline paper 12.1, 1999.

\bibitem{princstate} C.D. Poole and R.E. Wagner, ``Phenomenological approach to 
polarization dispersion in long single-mode fibers'', Electron. Letts.  22,
pp.1029-1030, 1986.


\bibitem{Matera90} F. Curti, B. Daino, G. de  Marchis  and  F.  Matera,  IEEE J. Lightwave Tech.
                {\bf 8}, 1162, 1990.

\bibitem{PMDcoherence} N. Gisin and JP. Pellaux, ``Polarization mode dispersion: time 
versus frequency domains'', Opt. Commun. 89, pp.316-323, 1992.

\bibitem{PMDCOST} N. Gisin and the COST 241 Group, {\it Definition of
polarization mode dispersion and first results of the COST 241 round-robin
measurements}, Pure and Applied Optics, {\bf 4}, 511-522, 1995.

\bibitem{Tsang} H.K. Tsang and R. Bhat, {\it Electron. Lett.} {\bf31}, 1773, 1995.

\bibitem{Reid} D.T. Reid et al., {\it Applied Optics} {\bf37}, 8142, 1998.

\bibitem{Scully99} M.O.Scully, B.-G. Englert and Ch.J. Bednar, {\it Phys. Rev. Lett.}
       {\bf83}, 4433, 1999.

\bibitem{Tomita} A. Tomita, "Complete Bell state measurement with controlled photon
       absorption and quantum interference", quant-ph/0006093.

\bibitem{InnsbruckSource} P. Kwiat et al., Phys. Rev. Lett. {\bf75}, 4337, 1995.

\bibitem{KwiatSource} P. Kwiat, E. Waks, A.G. White, I. Appelbaum and Ph. Eberhard,
       Phys. Rev. {\bf60A}, 773, 1999.

\bibitem{diffmode} That is, the two photons propagate in different directions, or
they propagate at different times (i.e. one behind the other).

\bibitem{optimal} A. Acin, R. Tarrach and G. Vidal, Phys. Rev. A {\bf61}, 62307, 2000.

\end{thebibliography}
\end{document}